\documentclass[journal, onecolumn, pagenumbers]{IEEEtran}
\usepackage{listings}
\usepackage{amsmath}
\usepackage{amsthm}
\usepackage{tikz}
\usepackage{caption}
\usepackage{array}
\usepackage{mdwmath}
\usepackage{multirow}
\usepackage{mdwtab}
\usepackage{eqparbox}
\usepackage{amsfonts}
\usepackage{tikz}
\usepackage{multirow,bigstrut,threeparttable}
\usepackage{amsthm}
\usepackage{array}
\usepackage{bbm}
\usepackage{subfigure}
\usepackage{epstopdf}
\usepackage{mdwmath}
\usepackage{mdwtab}
\usepackage{eqparbox}
\usepackage{tikz}
\usepackage{latexsym}
\usepackage{cite}
\usepackage{amssymb}
\usepackage{bm}
\usepackage{amssymb}
\usepackage{graphicx}
\usepackage{mathrsfs}
\usepackage{epsfig}
\usepackage{psfrag}
\usepackage{setspace}
\usepackage{hyperref}
\usepackage{algorithm}
\usepackage{algpseudocode}
\usepackage{stfloats}

\newtheorem{assumption}{Assumption}
\newtheorem{theorem}{Theorem}

\newtheorem{lemma}{Lemma} 
\newtheorem{corollary}{Corollary}

\newtheorem{definition}{Definition}

\def \cF {\mathcal{F}}

\def \bE {\mathbb{E}}

\def \bR {\mathbb{R}}

\def \cB {\mathcal{B}}

\def \cF {\mathcal{F}}

\def \bE {\mathbb{E}}
\def \bR {\mathbb{R}}
\def \bF {\mathbb{F}}

\def \cB {\mathcal{B}}

\newcommand{\Gscr}{{\cal G}}

\newcommand{\Pscr}{{\cal P}}

\newcommand\argmin{\mathop{\mbox{{\rm argmin}}}\limits}

\begin{document}

\title{Mutual Information, Relative Entropy and Estimation Error in Semi-martingale Channels}

\author{Jiantao~Jiao,~\IEEEmembership{Student Member,~IEEE},~Kartik~Venkat,~\IEEEmembership{Student Member,~IEEE}, and Tsachy~Weissman,~\IEEEmembership{Fellow,~IEEE}
\thanks{Jiantao Jiao, Kartik Venkat, and Tsachy Weissman are with the Department of Electrical Engineering, Stanford University, CA, USA. Email: \{jiantao, kvenkat, tsachy\}@stanford.edu. This work was supported in part by the Center for Science of Information (CSoI), an NSF Science and Technology Center, under grant agreement CCF-0939370. The material in this paper was presented in part at the 2016 IEEE International Symposium on Information Theory, Barcelona, Spain.}. 
}



%
\maketitle

\begin{abstract}
Fundamental relations between information and estimation have been established in the literature for the continuous-time Gaussian and Poisson channels, in a long line of work starting from the classical representation theorems by Duncan and Kabanov respectively. In this work, we demonstrate that such relations hold for a much larger family of continuous-time channels. We introduce the family of semi-martingale channels where the channel output is a semi-martingale stochastic process, and the channel input modulates the characteristics of the semi-martingale. For these channels, which includes as a special case the continuous time Gaussian and Poisson models, we establish new representations relating the mutual information between the channel input and output to an optimal causal filtering loss, thereby unifying and considerably extending results from the Gaussian and Poisson settings. Extensions to the setting of mismatched estimation are also presented where the relative entropy between the laws governing the output of the channel under two different input distributions is equal to the cumulative difference between the estimation loss incurred by using the mismatched and optimal causal filters respectively. The main tool underlying these results is the Doob--Meyer decomposition of a class of likelihood ratio sub-martingales. The results in this work can be viewed as the continuous-time analogues of recent generalizations for relations between information and estimation for discrete-time L\'evy channels. 
\end{abstract}


\section{Introduction}

The mutual information $I(X; Y)$ between two random objects $X, Y$ is defined as
\begin{equation}
I(X; Y) = \bE \log \frac{dP_{XY}}{d(P_X \times P_Y)}(X,Y),
\end{equation}
where the argument of the logarithm is the Radon--Nikodym derivative between the joint measure of $X$ and $Y$, and the product measure induced by $P_{XY}$. 

The mutual information $I(X;Y)$ plays a pivotal role in information theory, where it arises as the the maximal possible rate to communicate through a noisy channel defined by regular conditional probability distribution $P_{Y|X}$~\cite{Shannon1949}. This paper deals with the characterization of mutual information under general observation models involving continuous-time stochastic processes. This problem has a rich history. Duncan~\cite{Duncan1970} considered the problem of explicitly characterizing the mutual information in the setting of the canonical white Gaussian channel. Under this channel model, the output process $\{Y_t: 0\leq t\leq T\}$ satisfies the following stochastic differential equation:
\begin{equation}\label{eqn.awgn}
dY_t = \sqrt{\gamma} X_t dt + dW_t,
\end{equation}
where the input process $ X^T = \{X_t: 0\leq t \leq T \}$ is independent of the standard Brownian motion $W^T = \{W_t: 0\leq t\leq T\}$, and $\gamma$ is the Signal-to-Noise-Ratio(SNR) parameter. In cases where we need to explicitly show the SNR level, we denote the random variable $Y_t$ as $Y_{\gamma,t}$, and the whole process $Y^T$ as $Y_\gamma^T$. Duncan~\cite{Duncan1970} showed that if the channel input $X_t$ satisfies a finite power constraint, then the mutual information takes the following form:
\begin{equation}\label{eqn.duncan}
I(X^T; Y^T) = \frac{\gamma}{2}\int_0^T \bE (X_t - \bE[X_t|Y^t])^2 dt.
\end{equation}

Equation~(\ref{eqn.duncan}) is remarkable since it obtains an explicit formula for the mutual information, for essentially any input process corrupted by white Gaussian noise. Further, it reveals an intimate connection between the mutual information and the \emph{minimum mean squared error} in estimating the channel input $X_t$ based \emph{causally} on the output process $Y_t$. For instance, this result provides the insight that the capacity achieving input distribution which maximizes the mutual information, must also be the one that is hardest to estimate under squared loss. The rich interconnections between information measures and the corresponding loss incurred in estimation are one of the central themes of this work. Duncan's result~\eqref{eqn.duncan} is the first of many important milestones for relations between information and estimation in continous-time channels. 

Kadota, Ziv and Zakai~\cite{Kadota--Zakai--Ziv1971} extended the relation above to the continuous-time white Gaussian channel in the presence of causal feedback. They proved that
\begin{equation}\label{eqn.kzz}
I(\alpha; Y^T) = \frac{\gamma}{2} \int_0^T \bE(X_t(\alpha,Y^t) - \bE[X_t(\alpha,Y^t)|Y^t])^2 dt,
\end{equation}
where $\alpha$ is the continuous-time message process to be transmitted, and the channel input $X_t(\alpha,Y^t)$ which encodes the message, depends causally on the output process $Y_t$ and the message $\alpha$. 

This relationship has immediate implications. For example,~\cite{Kadota--Zakai--Ziv1971} used (\ref{eqn.kzz}) to show that feedback does not increase the capacity of continuous-time white Gaussian channel. It is worth noting that the channel without feedback is subsumed in the case with feedback if we take $\alpha = X^T$, i.e. the channel input is the message itself. From now on we will consider the  \emph{more general} case where feedback is allowed.

Paralleling the developments in the white Gaussian channel, in 1978 Kabanov~\cite{Kabanov1978} calculated the capacity for continuous-time Poisson channel with feedback. Suppose the output process $Y^T = \{Y_t, 0\leq t\leq T\}$ is a point process whose compensator (stochastic intensity) is $\gamma \int_0^t X_s ds$, where $X_t = X_t(\alpha,Y^{t-})$ is the \emph{predictable} input process, and $\alpha$ is the message. This is the so-called {\em continuous-time Poisson channel with feedback}. Adopting notations introduced in \cite{Atar--Weissman2012}, we know from~\cite[Thm. 19.11.]{Liptser--Shiryaev2001}\cite{Guo--Shamai--Verdu2008} that if $\int_0^T \bE X_t \log X_t dt <\infty$, then
\begin{equation}
I(\alpha,Y^T) = \gamma \int_0^T \bE \ell_\Pscr(X_t, \bE[X_t|Y^{t-}])dt,
\end{equation} 
where $\ell_\Pscr(x,y) = x \ln(x/y) - x+ y, x>0, y>0$ is the \emph{natural} loss function for estimation in the Poisson channel. 

Our main contribution in this work is to introduce a class of semi-martingale channels and to present a new formula for the mutual information in the same spirit as the relations above for the Gaussian and Poisson channels. In particular, the family of semi-martingale channels will include the continuous-time Gaussian and Poisson channels as special cases, and the new formula for mutual information under this model will generalize and unify the two classical results presented above, as well as present new relations between information and estimation.  We note that generalized representations of mutual information are a topic of great interest, and recent efforts in that direction include~\cite{Duncan2010}, which presents estimation theoretic formulae for mutual information between a stochastic signal and a pure jump L\'evy process which is modulated by the signal, and~\cite{Johnson2013} where a generalization of the famous de Bruijn's identity is presented for general families of stable densities. Beyond the Gaussian and Poisson models, \cite{Grigelionis1974} calculated the mutual information for locally infinitely divisible processes in 1974. 

As part of the history of results discovered for the continuous-time Gaussian and Poisson channels, we include here some of the more recent developments and insights which are informed by relations between information and estimation. After recapping these extensions, we will introduce the framework for results in this paper. 
\subsubsection{Deriving scalar channel results from continuous-time families}
Before proceeding to develop generalizations for continuous-time families, we quickly recap the scalar Gaussian channel and the I-MMSE relationship~\cite{Guo--Shamai--Verdu2005} which presents the derivative of the mutual information (with respect to SNR) as the minimum mean squared error in estimation of the channel input based on the noisy observation. We can re-write the scalar I-MMSE as:  
\begin{equation}\label{eqn.i-mmse}
\frac{\partial}{\partial \gamma} I(X; \sqrt{\gamma} X + N) = \frac{1}{2} \bE (X - \bE[X|\sqrt{\gamma}X + N])^2,
\end{equation}
where $\bE X^2 <\infty, N\sim \mathcal{N}(0,1)$, $X$ is independent of $N$, and $\gamma > 0$. Among its many applications include proving the entropy power inequality in \cite{Verdu2006}, and the monotonic decrease of the non-Gaussianness of the sum of independent random variables in \cite{Tulino--Verdu2006}. 

It is worth noting that the I-MMSE relationship~\eqref{eqn.i-mmse} can be directly obtained as a corollary to Duncan's theorem~\eqref{eqn.duncan}. Indeed, if we take $Y_\gamma = \gamma X + W_\gamma$, $W_\gamma$ a standard Brownian motion indexed by $\gamma \geq 0$, then by Duncan's theorem we know that
\begin{equation}\label{eqn.duncanimply}
I(X; Y_\gamma) = \frac{1}{2}\int_0^\gamma \bE (X - \bE[X|Y_\alpha])^2 d\alpha,
\end{equation}
where we have used the fact that $Y_\gamma$ is the sufficient statistic for parameter $X$ given $\{Y_\alpha\}_{0\leq \alpha \leq \gamma}$. Taking derivative with respect to $\gamma$ on both sides of (\ref{eqn.duncanimply}), we arrive at the I-MMSE relationship. Analogously, results paralleling I-MMSE in the Poisson channel settings appear in \cite{Guo--Shamai--Verdu2008}, \cite{Atar--Weissman2012}, where again they can be shown to be corollaries of the (more general) results for the continuous-time Poisson channel. 

\subsubsection{Extensions to mismatched estimation and relative entropy}
Recall that the relative entropy $D(P \| Q)$, is defined between two probability measures $P \ll Q$, as follows
\begin{equation}
D(P \| Q) = \bE_P \log \frac{dP}{dQ}.
\end{equation}

We emphasize that the I-MMSE relations can be recovered from the results of mismatched estimation. Indeed, we have
\begin{align}
I(X;Y) & = \mathbb{E} D (P_{Y|X} \| P_Y),
\end{align}
and $P_{Y|X}$ can be viewed as the output distribution of a channel with deterministic input $X$, and $P_Y$ can be viewed as the marginal output distribution. 

Weissman~\cite{Weissman2010} presented a representation formula for relative entropy in continuous-time white Gaussian channels with feedback. Let $P$ and $Q$ denote two probability measures on the input process $X^T$, and the channel model is the same as in (\ref{eqn.awgn}). Under mild conditions, the main result of~\cite{Weissman2010} shows that
\begin{equation}
D(P_{Y_\gamma^T} \| Q_{Y_\gamma^T}) = \frac{\gamma}{2} \left( \mathsf{cmse}_{P,Q}(\gamma) - \mathsf{cmse}_{P,P}(\gamma)  \right),
\end{equation}
where $\mathsf{cmse}_{P,Q}(\gamma) = \int_0^T \bE_P(X_t - \bE_Q[X_t|Y^t])^2 dt$ denotes the mismatched filtering error under squared error loss. The paralleling mismatched estimation interpretations of relative entropy in the Poisson channel settings was demonstrated in~\cite{Atar--Weissman2012}.

\subsubsection{Pointwise extensions}
\cite{Venkat--Weissman2012} and \cite{Jiao--Venkat--Weissman2013} showed a pointwise analog of the relations above in the Gaussian and Poisson settings, respectively. One particular feature of these results is the Doob--Meyer decomposition of a class of sub-martingales, i.e. the $P$-sub-martingales
\begin{equation}\label{eqn.density}
\log \frac{dP_{Y^t}}{dQ_{Y^t}}, \log \frac{dP_{Y^t|\alpha}}{dP_{Y^t}}
\end{equation}
where $Y_t$ is the output process of a continuous-time white Gaussian channel or a Poisson channel. Conceivably, the predictable non-decreasing part of their Doob--Meyer decomposition corresponds to an estimation error term, and the local martingale part corresponds to a stochastic integral. The results corresponding to relative entropy can be obtained by taking expectations of these sub-martingales.

Having revisited the rich historical results in continuous-time channels, a natural question arises: do Gaussian and Poisson models capture the whole picture relations between information and estimation? Do there exist natural extensions of the results above beyond Gaussian and Poisson models which preserve the estimation-theoretic interpretations for important information measures? The authors answered this question affirmatively for scalar transformations by defining the general class of discrete-time L\'evy channels~\cite{Jiao--Venkat--Weissman_ISIT2014,Jiao--Venkat--Weissman_TIT2014}. In this paper, we show that the answer is affirmative for continuous-time channels. Concretely, our contributions in this spirit span the following aspects:
\begin{enumerate}
\item We propose a general definition of \emph{semi-martingale} channels, which includes as special cases, the white Gaussian channel, and the Poisson point process channel. 
\item For semi-martingale channels, we obtain the input-output mutual information as the minimum causal estimation error under a natural loss function, thereby extending the findings for Gaussian and Poisson channels in continuous-time. 
\item We also extend the above result to the setting of mismatched estimation and obtain a new representation for the relative entropy as the cost of mismatch in estimation under the same loss function for semi-martingale channels.
\item We also obtain pointwise extensions for these identities via expressions for sub-martingales in (\ref{eqn.density}) when $Y_t$ is the output of a general semi-martingale channel;
\end{enumerate}

We note that this work can be viewed as the continuous-time analog of~\cite{Jiao--Venkat--Weissman_TIT2014}, where the authors introduce discrete-time L\'evy channels. The rest of this paper is organized as follows. Section \ref{sec: SMG prelim} will review some preliminaries. We will present the main results on continuous-time semi-martingale channels in Section~\ref{sec: SMG channels}. We then discuss the main proof elements in Section~\ref{sec: SMG proofs}, and present our conclusions in Section~\ref{sec: SMG summary}.

\section{Preliminaries} \label{sec: SMG prelim}

\subsection{Semi-martingales}

We assume as given a complete probability space $(\Omega,\cF,P)$. In addition we are given a filtration $(\cF_t)_{0\leq t\leq \infty}$. By a filtration we mean a family of $\sigma$-algebras $(\cF_t)_{0\leq t\leq \infty}$ that is increasing, i.e., $\cF_s \subset \cF_t$ if $s\leq t$. For convenience, we will usually write $\bF$ for the filtration $(\cF_t)_{0\leq t\leq \infty}$. We denote $\cF_t^Y = \sigma\{Y_s: s\leq t\}$ to be the natural filtration generated by stochastic process $Y$, and $\sigma\{X_\zeta, \zeta \in \mathcal{Z}\}$ denotes the smallest $\sigma$-algebra with respect to which $X_\zeta$ is measurable. We have $\mathcal{F}_{t-} = \sigma \left(  \bigcup_{s<t} \mathcal{F}_s \right)$. 

By $D[0,T]$ we denote the space of real-valued functions $y(t)$ defined on $[0,T]$ which are cadlag, i.e., right-continuous with left limits. We also denote the space of real-valued continuous functions $y(t)$ on $[0,T]$ by $C[0,T]$. Note that here $T$ could be taken to be $\infty$, in that case, the interval $[0,T]$ should be interpreted as $[0,\infty)$. We equip the space $D[0,T]$ with Skorokhod topology, and the space $C[0,T]$ with sup-norm topology. We define the Borel $\sigma$-algebras $\cB_t(C) = \sigma\{y_s, s\leq t, y \in C[0,T]\}$ and $\cB_t(D) =\sigma\{y_s, s\leq t, y \in D[0,T]\}$. 

For simplicity, throughout this paper, we only deal with one-dimensional real-valued stochastic processes. However it is worth noting that our results can be easily generalized to higher dimensions. 

There exist various version of definitions for semi-martingales, and we adopt the following version.
\begin{definition}\cite[Def. 2.17]{medvegyev2007stochastic}\label{def.semimartingale}
An adapted process $X$ is called a semi-martingale if $X$ has a decomposition
\begin{align}
X = X_0 + V + H,
\end{align}
where $V$ is a right-continuous, adapted processes with finite variation, $H$ is locally square integrable, and $V_0 = H_0 = 0$. 
\end{definition}

The class of semi-martingales is a very broad one. Indeed, it consists of every local martingale, and every integrable sub-martingale and super-martingales. For continuous semi-martingales the decomposition in Definition~\ref{def.semimartingale} is unique~\cite[Prop. 2.19]{medvegyev2007stochastic}. 

It is well known~\cite[Chap. 4.1]{Liptser--Shiryaev1989} that any cadlag semi-martingale $Y_t$ can be represented as
\begin{equation}
Y_t = Y_0 + B_t + Y_t^c + \int_0^t \int_{|z|\leq 1} z d(\mu - \nu) + \int_0^t \int_{|z|>1} z d\mu,
\end{equation}
where $B$ is a predictable process of locally bounded variation, $B_0 = 0$; $Y^c$ is the continuous local martingale component of the semi-martingale $Y$; $\mu$ is the jump measure of $Y$, and $\nu$ is its compensator. The jump measure $\mu = \mu(dt, dz)$ has two arguments, which satisfies the following relation:
\begin{equation}
\mu((0,t]\times \Gamma) = \sum_{0<s\leq t} I(\Delta Y_s \in \Gamma), \Gamma \in \cB(\bR_0), \bR_0 = \bR \backslash \{0\},
\end{equation}
where $\cB(\bR_0)$ is the Borel $\sigma$-algebra on $\bR_0$. Informally, $\mu(dt,dz)$ counts the number of jumps of size $z$ at time $t$, and its compensator $\nu(dt,dz)$ characterizes the intensity of jumps of size $z$ at time $t$. 

For simplicity, we assume $\nu(\{t\} \times \bR_0) = 0, \forall t \geq 0$. That is to say, $\nu((0,t]\times \bR_0)$ is a continuous function of $t$. Let $C = [Y^c, Y^c]_t$ be the quadratic variation process of $Y^c$. The quadratic variation process of the continuous stochastic process $Y^c$ is defined as 
\begin{align}
[Y^c, Y^c]_t & =\lim _{\Vert m\Vert \rightarrow 0}\sum _{k=1}^{n}(Y^c_{t_{k}}-Y^c_{t_{k-1}})^{2},
\end{align}
where $m$ ranges over partitions of the interval $[0,t]$ and the norm of the partition $m$ is the mesh $\max\{(t_i - t_{i-1}): i = 1,2,\ldots,n\}$. The limit, if it exists, is defined using convergence in probability. We call collection $(B,C,\nu)$ the \emph{triplet of predictable characteristics} of a semi-martingale $Y$. The triplet is uniquely determined by the process $Y$.

In general, unfortunately, the triplet does \emph{not} fully specify the distribution of the semi-martingale $Y$ (cf. Example 1.9 of~\cite{Rao1999}). Hence, to avoid some unnecessary technical difficulties, throughout this paper, we assume all semi-martingales satisfy the property of $(\tau_n)$-uniqueness (also called \emph{local uniqueness} in the literature~\cite[Pg. 159]{Jacod--Shiryaev1987}), which is defined as follows:

\begin{definition}[$\tau_n$ uniqueness]\cite[Sec. 11]{Kabanov--Liptser--Shiryaev1978-2} \label{def.taununiqueness}
The measure $P$ of a semi-martingale $Y$ is said to have the property of $(\tau_n)$-uniqueness if the triplets $(B^{\tau_n}, C^{\tau_n}, \nu^{\tau_n})$ of process $Y_{t\wedge \tau_n}$ uniquely determine the restrictions $P_{\tau_n}$ of the measure $P$ to the $\sigma$-algebras $\cF_{\tau_n}$. Here $\tau_n$ is any sequence of $\cF_t$-stopping times such that $\tau_n \uparrow \infty, P$-a.s.
\end{definition}

The $(\tau_n)$-uniqueness property was first introduced in \cite{Jacod--Memin1976}, and has been established so far for semi-martingales with independent increments, diffusion type processes, multivariate point processes in~\cite{Jacod--Shiryaev1987}, and for Markov processes in \cite{Skorokhod1982} and \cite{Kabanov--Liptser--Shiryaev1980}.

\subsection{L\'evy processes and Infinitely divisible distributions}

A general one-dimensional L\'evy process is defined as follows.
\begin{definition}[L\'evy process]
A process $Y = \{Y_t: t\geq 0\}$ defined on a probability space $(\Omega, \mathcal{F}, \mathbb{P})$ is said to be a L\'evy process if it possesses the following properties:
\begin{enumerate}
\item The paths of $Y$ are $\mathbb{P}$-almost surely right continuous with left limits.
\item $\mathbb{P}(Y_0 = 0) = 1$. 
\item For $0\leq s\leq t$, $Y_t - Y_s$ is equal in distribution to $Y_{t-s}$. 
\item For $0\leq s\leq t, Y_t - Y_s$ is independent of $\{Y_u: u\leq s\}$. 
\end{enumerate}
\end{definition}

L\'evy processes belong to the class of semi-martingales, where its predictable characteristics are non-random and the $(\tau_n)$-uniqueness property is satisfied. Important examples of L\'evy processes include include Brownian motion and Poisson processes. We refer the reader to Sato \cite{Sato1999} for a comprehensive treatment of L\'evy processes.

The infinitely divisible distribution is defined as follows:
\begin{definition}[Infinitely divisible distributions]
We say that a real-valued random variable $T$ has an infinitely divisible distribution if for each $n \in \mathbb{N}, n\geq 1$, there exists a sequence of i.i.d. random variables $T_{1,n}, T_{2,n},\ldots, T_{n,n}$ such that
\begin{align}
T \stackrel{d}{=} T_{1,n} + T_{2,n} + \ldots + T_{n,n},
\end{align}
where $\stackrel{d}{=}$ is equality in distribution.  
\end{definition}

The Gaussian, Poisson, negative binomial, gamma and Cauchy distributions are all infinitely divisible distributions on $\Re$. 

From the definition of a L\'evy process we see that for any $t>0$, $Y_t$ is a random variable belonging to the class of infinitely divisible distributions. Indeed, it follows from the fact that for any $n = 1,2,\ldots$, 
\begin{align}
Y_t = Y_{t/n} + (Y_{2t/n} - Y_{t/n}) + \ldots + (Y_t - Y_{(n-1)t/n})
\end{align}
together with the fact that $\{Y_t\}$ has stationary independent increments. 

The following lemma relates the characteristic exponent of $Y_t$ with that of $Y_1$. 
\begin{lemma}\cite[Chap. 2.1.]{Kuchler--Sorensen1997}\label{lemma.levycumulant}
For a L\'evy process $Y_t$, if $\mathbb{E} e^{i\theta Y_t} = e^{ \Psi_t(\theta)}$, then $\Psi_t(\theta) = t \Psi_1(\theta)$. 
\end{lemma}

Indeed, for two positive integers we have
\begin{align}
m \Psi_1(\theta)  = \Psi_m(\theta) = n \Psi_{m/n}(\theta),
\end{align}
which proves the statement for all rational $t>0$. The irrational cases follows from taking a limit and applying the right continuity of $X_t$ and the dominated convergence theorem. 

The full extent to which we may characterize infinitely divisible distributions is described by the L\'evy--Khintchine formula. 
\begin{lemma}[L\'evy--Khintchine formula]\cite{Sato1999}\label{lemma.lemmakhintchine}
A real-valued random variable $Y$ is infinitely divisible with characteristic function represented as
\begin{align}
\mathbb{E} e^{i \theta Y} & = e^{\Psi(\theta)},\quad \theta \in \mathbb{R},
\end{align}
if and only if there exists a triple $(a, \sigma, \nu)$, where $a\in \mathbb{R}, \sigma\geq 0$, and $\nu(\cdot)$ is a measure concentrated on $\mathbb{R} \backslash \{0\}$ satisfying $\int_{\mathbb{R}} (1 \wedge x^2) \nu(dx) <\infty$, such that
\begin{align}
\Psi(\theta) & = i a \theta - \frac{1}{2} \sigma^2 \theta^2 + \int_{\mathbb{R}} (e^{ i \theta z}-1 - i\theta z \mathbbm{1}_{|z|<1})\nu(dz). 
\end{align}
\end{lemma}

We call the tuple $(a,\sigma,\nu(dz))$ \emph{L\'evy characteristics} of the L\'evy process $\{Y_t\}$ if the characteristic function of $Y_1$ follows the L\'evy--Khintchine formula with triplet $(a,\sigma,\nu(dz))$. Particularly, we call the number $\sigma$ \emph{diffusion coefficient}, and the measure $\nu(dz)$ the \emph{L\'evy measure} of the L\'evy process $\{Y_t\}$. 

We have seen so far, that every L\'evy process can be associated with the law of an infinitely divisible distribution. The opposite, i.e. that given any random variable $X$, whose law of infinitely divisible, we can construct a L\'evy process $\{Y_t\}$ such that $Y_1 \stackrel{d}{=}X$. This is the subject of the L\'evy--It$\hat{\mathrm{o}}$ decomposition. 
\begin{lemma}\cite[Chap. 4]{Sato1999}[L\'evy--It$\hat{\mathrm{o}}$ decomposition]\label{lemma.levyitodecomposition}
Consider a triplet $(a,\sigma,\nu)$ where $a\in \mathbb{R}, \sigma\geq 0$ and $\nu$ is a measure satisfying $\nu(\{0\}) = 0$ and $\int_{\mathbb{R}} (1 \wedge x^2) \nu(dx)<\infty$. Then, there exists a probability space $(\Omega, \mathcal{F}, \mathbb{P})$ on which a L\'evy process $\{Y_t\}$ exists and decomposes as four independent processes as
\begin{align}\label{eqn.levyito}
Y_t = at + \sigma W_t + \int_0^t \int_{|z|< 1} z (\mu(ds,dz) - \nu(dz)ds) + \int_0^t \int_{|z| \geq 1} z \mu(ds,dz),
\end{align}
where $W_t$ is a standard Brownian motion, $\int_0^t \int_{|z|< 1} z (\mu(ds,dz) - \nu(dz)ds)$ is a square integrable pure jump martingale with an almost surely countable number of jumps of magnitude less than one on each finite time interval, and $\int_0^t \int_{|z| \geq 1} z \mu(ds,dz)$ is a compound Poisson process. The $\mu(dt,dz)$ is a jump measure defined to satisfy the following relations: $\forall \, \Gamma\in \mathcal{B}(\mathbb{R} \backslash \{0\})$,
\begin{equation}\label{eqn.jumpmeasuredef}
\mu((0,t] \times \Gamma) = \sum_{0<s\leq t} \mathbb{I}(\Delta Y_s \in \Gamma), 
\end{equation}
where $ \Delta Y_s = Y_s - Y_{s-}, Y_{s-} = \lim_{u\to s-} Y_u$. The measure $\nu(dz)$ is defined such that
\begin{equation}
\int_0^t \int_{|z|< 1} z (d\mu - \nu(dz)ds)
\end{equation}
is a martingale indexed by $t$. The measure $\nu(dz)ds$ is called the compensator for the multivariate point process $\mu(ds,dz)$. 
\end{lemma}

\subsection{Semi-Martingale Channels}

We assume, when there is no input signal, the channel output is a L\'evy process. We assume the SNR level is $\gamma$. By the L\'evy-It$\hat{\mathrm{o}}$ decomposition in Lemma~\ref{lemma.levyitodecomposition}, given any L\'evy process $Y_t$, there exist constants $a \in \bR,\sigma\geq 0$, a non-negative measure $\nu(\cdot)$ on $\cB(\bR_0)$ s.t. $\int_{\bR_0} \min(1,z^2)\nu(dz)<\infty$, such that the predictable characteristics of $Y$ satisfy
\begin{equation}
B_t = at, C_t = \sigma t, \nu(dt,dz) = \gamma \nu(dz)dt.
\end{equation}

In order to be consistent with results for Gaussian and Poisson channels, without loss of generality in this section we take $a = 0, \sigma = 1$. That is to say, in the absence of input signal, the output process $(Y_t, \bF, P_0)$ of a semi-martingale channel at SNR $\gamma$ is a L\'evy process with the following representation:
\begin{equation}\label{eqn.levyitoSMG}
Y_t = W_t + \int_0^t \int_{|z|< 1} z (d\mu - \gamma \nu(dz)ds) + \int_0^t \int_{|z|\geq 1} zd \mu,
\end{equation}
where $W_t$ is a standard Brownian motion, $\mu(dt,dz)$ is a Poisson random measure on $[0,T]\times \bR_0$, independent of $W^T$.  

Now we specify the output given message $\alpha$. We assume the message $\alpha$ takes values in a measurable space $(A, \mathcal{A})$. For any $s\geq 0$, let $\beta_s = \beta_s(\alpha, Y^{s-})$ be a $\mathcal{A} \otimes \cB_{s-}(D)$-measurable function. For any $s\geq 0, z\in \bR_0$, let $\lambda_{s,z} = \lambda_{s,z}(\alpha,Y^{s-})\geq 0$ also be a $\mathcal{A} \otimes \cB_{s-}(D)$-measurable function. The functions $\beta_s(\alpha, Y^{s-}), \lambda_{s,z}(\alpha,Y^{s-})$ are called \emph{encodings} of $\alpha$ for transmission over the semi-martingale channel. At SNR level $\gamma$, the output $(Y, \bF, P)$ corresponding to a semi-martingale channel with encodings $\beta_s(\alpha, Y^{s-}), \lambda_{s,z}(\alpha,Y^{s-})$ satisfies the following representation:
\begin{align}
Y_t & = \sqrt{\gamma}\int_0^t \beta_s ds +  W_t + \gamma \int_0^t \int_{|z|< 1} z (\lambda_{s,z}-1) \nu(dz)ds  \nonumber \\
& \quad + \int_0^t \int_{|z|< 1} z (d\mu - \gamma \lambda_{s,z}\nu(dz)ds) + \int_0^t \int_{|z|\geq 1} z d\mu,
\end{align}
where $W_t$ is a standard Brownian motion under $P$. In other words, the predictable characteristics of the output process $Y$ has changed from $(0,t,\gamma \nu(dz)dt)$ to 
\begin{equation}\label{eqn.pcnew}
(\sqrt{\gamma}\int_0^t \beta_s ds + \int_0^t \int_{|z|< 1} \gamma z (\lambda_{s,z}-1)\nu(dz)ds, t, \gamma \lambda_{t,z} \nu(dz)dt).
\end{equation}
The $(\tau_n)$-uniqueness property guarantees that the distribution of the output process $Y_t$ is uniquely determined by the input signals $\beta_s(\alpha, Y^{s-})$ and $\lambda_{s,z}(\alpha, Y^{s-})$. 

Note that the definition of the semi-martingale channel generalizes those of the white Gaussian and Poisson channels. Indeed, the semi-martingale channel degenerates to the white Gaussian channel when $\nu(dz) \equiv 0$, and it degenerates to the Poisson channel when $\nu(dz) = \delta_{z = 1}$, $\beta_s \equiv 0$ and the Brownian motion part disappear. 

Throughout this paper, we assume the following conditions. 
\begin{assumption}\label{ass.inputdefinition}
We assume the following throughout this paper:
\begin{enumerate}
\item Any filtered complete probability space $(\Omega,\cF, \bF, P)$ satisfies the \emph{usual hypotheses}, i.e.
\begin{enumerate}
\item $\cF_0$ contains all the $P$-null sets of $\cF$;
\item $\cF_t = \bigcap_{u>t} \cF_u, \forall t, 0\leq t<\infty$; that is, the filtration $\cF$ is right-continuous. 
\end{enumerate}
\item All the processes satisfy the $(\tau_n)$-uniqueness property as defined in Definition~\ref{def.taununiqueness}. 
\item There exists a constant $V>0$ such that with probability one,
\begin{align}
\int_0^T \beta_s^2 ds + \int_0^T \int_{\mathbb{R}_0} (1-\sqrt{\lambda_{s,z}})^2 \nu(dz) ds \leq V. 
\end{align}
\item $ \int_0^T \mathbb{E}|\beta_s|ds < \infty$, $\int_0^T \int_{\mathbb{R}_0} \mathbb{E}|\lambda_{s,z}| \nu(dz) ds< \infty$. 
\item For any $0\leq s\leq T$, $\mathbb{E}|\beta_s| <\infty$, $ \mathbb{E} \int_{\mathbb{R}_0} \lambda_{s,z} \nu(dz) < \infty$. 
\end{enumerate}
\end{assumption}

We emphasize that the conditions in Assumption~\ref{ass.inputdefinition} allows us to avoid messy and delicate measure theoretic details related to the definition of predictable projections and predictable $\sigma$-algebras.

\section{Main results} \label{sec: SMG channels}

\subsection{Calculation of the Mutual Information}

Now we state a theorem on representation of the mutual information $I(\alpha; Y^T)$ in the semi-martingale channel, which is the main result of this paper.
\begin{theorem}\label{thm.mutual}
Under Assumption~\ref{ass.inputdefinition}, if 
\begin{align}
\int_0^T \mathbb{E} \beta_s^2 ds + \int_0^T \mathbb{E} \lambda_{s,z} \left|\ln  \frac{\lambda_{s,z}}{\hat{\lambda}_{s,z}^{P,C}(\gamma)} \right | \nu(dz) ds <\infty,
\end{align}
then,
\begin{equation}\label{eqn.thm1}
I(\alpha;Y^T) = \gamma \left [ \int_0^T \bE \ell_\Gscr(\beta_s ,\hat{\beta}_{s}^{P,C}(\gamma))ds + \int_0^T \int_{\bR_0} \bE \ell_\Pscr(\lambda_{s,z}, \hat{\lambda}_{s,z}^{P,C}(\gamma)) \nu(dz)ds \right],
\end{equation}
where $\hat{\beta}_s^{P,C}(\gamma) = \mathbb{E}_P[\beta_s|\cF_{s-}^Y]$, $\hat{\lambda}_{s,z}^{P,C}(\gamma) = \mathbb{E}_P[\lambda_{s,z}| \cF_{s-}^Y]$. The loss functions $\ell_\Gscr(x,y) = \frac{1}{2}(x-y)^2$, $\ell_\Pscr(x,y) = x\ln(x/y)-x + y$.
\end{theorem}

Here we need to explain the notation a little. The superscripts $P$ and $C$ in notations $\hat{\beta}_s^{P,C}(\gamma)$ and $\hat{\lambda}_{s,z}^{P,C}(\gamma)$ mark the fact that the conditional expectation is taken under probability law $P$ \emph{Causally} on the history of $Y$. We emphasize that both the loss functions $\ell_\Gscr$ and $\ell_\Pscr$ are Bregman divergences. We introduce the notion of the Bregman divergence below. 
\begin{definition}\label{def.bregmandivergence}
Let $f: \Omega \mapsto \mathbb{R}$ be a convex, continuously differentiable function, the domain $\Omega \subset \mathbb{R}^d$. Then, the Bregman divergence associated with $f$, denoted as $d_f(x,y)$, is defined as
\begin{align}
d_f(x,y) = f(x) - f(y) - \langle \nabla f(y), x-y \rangle,
\end{align}
where $\langle x,y \rangle$ denotes the inner product of $x$ and $y$. 
\end{definition}

It follows from Jensen's inequality that $d_f(x,y)\geq 0$. It is clear that $\ell_{\mathcal{G}}(x,y) = d_f(x,y)$ when $f = \frac{1}{2} x^2$, and $\ell_{\mathcal{P}}(x,y) = d_f(x,y)$ when $f = x\ln x$. The Bregman divergence satisfies the following property when used as a loss function in Bayesian decision theory:
\begin{lemma}\label{lemma.meanpropertybregman}
Suppose $X$ is a random variable taking values in $\Omega$. Then, for any non-random element $u\in \Omega$, 
\begin{align}
\mathbb{E}[d_f(X,u)] & = \mathbb{E}[d_f(X, \mathbb{E}[X])] + d_f(\mathbb{E}[X],u),
\end{align}
where the expectations are taken with respect to the distribution of $X$. 
\end{lemma}

\begin{IEEEproof}
It follows from straightforward algebra that
\begin{align}
d_f(X,u) & = d_f(X, \mathbb{E}[X]) + d_f(\mathbb{E}[X],u) + \langle f'(\mathbb{E}[X]) - f'(u), X - \mathbb{E}[X] \rangle. 
\end{align}
Taking expectations on both sides finishes the proof. 
\end{IEEEproof}

It follows from Lemma~\ref{lemma.meanpropertybregman} that
\begin{align}
\mathbb{E}[X] & = \argmin_{u\in \Omega} \mathbb{E}[d_f(X,u)]. 
\end{align}
Further, if $f$ is strictly convex, then $\mathbb{E}[X]$ uniquely solves $\min_u \mathbb{E}[d_f(X,u)]$. It is sometimes called the \emph{orthogonality principle}.

\subsection{Relative entropy representations}

Assume $P$ and $Q$ are two probability measures on the inputs $(\beta_, \lambda_{s,z})$ to the semi-martingale channel. We denote the \emph{mismatched causal estimation error} at SNR $\gamma$ as
\begin{equation}
\mathsf{cmle}_{P,Q}(\gamma) = \int_0^T \bE_P \ell_\Gscr(\beta_s , \hat{\beta}_s^{Q,C}(\gamma)) ds + \int_0^T \int_{\bR_0} \bE_P \ell_\Pscr(\lambda_{s,z}, \hat{\lambda}_{s,z}^{Q,C}) \nu(dz)ds,
\end{equation}
where $\hat{\beta}_s^{Q,C}(\gamma) = \bE_Q[\beta_s|\cF_{s-}^Y], \hat{\lambda}_{s,z}^{Q,C} = \bE_Q[\lambda_{s,z}|\cF_{s-}^Y]$. 

According to Theorem~\ref{thm.mutual}, we know
\begin{equation}
I(\alpha; Y^T) = \gamma \cdot \mathsf{cmle}_{P,P}(\gamma). 
\end{equation}

A natural interpretation of the quantity
\begin{equation}
\mathsf{cmle}_{P,Q}(\gamma) - \mathsf{cmle}_{P,P}(\gamma)
\end{equation}
is the \emph{penalty of mismatch} in estimation under probability measure $P$. In other words, it is the \emph{excessive} estimation error caused by the fact that the decoder takes the distribution of the inputs as $Q$ while the true distribution is $P$. By the orthogonality principle of $\ell_\Gscr$ and $\ell_\Pscr$, we know it is never negative, and intuitively it could serve as a measure quantifying the distance between probability measures $P$ and $Q$. This intuition is rigorized by the following theorem.

\begin{theorem}\label{thm.relative}
Under Assumption~\ref{ass.inputdefinition}, if
\begin{align}
\int_0^T \mathbb{E}\left( \hat{\beta}_s^{P,C}(\gamma) - \hat{\beta}_s^{Q,C}(\gamma)\right)^2 ds + \int_0^T \mathbb{E} \lambda_{s,z} \left|  \ln \frac{\hat{\lambda}_{s,z}^{P,C}(\gamma)}{\hat{\lambda}_{s,z}^{Q,C}(\gamma)} \right | \nu(dz) ds <\infty,
\end{align}
then
\begin{equation}
D(P_{Y^T_\gamma} \| Q_{Y^T_\gamma}) = \gamma \cdot \left( \mathsf{cmle}_{P,Q}(\gamma) - \mathsf{cmle}_{P,P}(\gamma) \right),
\end{equation}
where $\hat{\beta}_s^{Q,C}(\gamma) = \bE_Q[\beta_s|\cF_{s-}^Y], \hat{\lambda}_{s,z}^{Q,C} = \bE_Q[\lambda_{s,z}|\cF_{s-}^Y]$. 
\end{theorem}

\subsection{Special Cases: White Gaussian Channels and Multivariate Point Process Channels} \label{sec.special}

We emphasize that for special classes of the semi-martingale channel, such as the AWGN channel and the multivariate point process channel, we can obtain similar results under much weaker conditions on the input processes. 

\subsubsection{White Gaussian Channel}
First we deal with the white Gaussian channel. As proved in \cite{Jacod--Shiryaev1987}, the $(\tau_n)$-uniqueness property is satisfied in this case. In fact in this case we can considerably weaken the assumptions to~\cite[Chap. 16.3]{Liptser--Shiryaev2000}
\begin{equation}\label{eqn.condg}
\int_0^T \bE \beta_s^2 ds<\infty,
\end{equation}
which has the natural interpretation of restricting the total power of input signals. Under~(\ref{eqn.condg}), we have the classic result by \cite{Kadota--Zakai--Ziv1971}:

\begin{corollary}\label{cor.kzz}
Under channel model
\begin{equation}
d Y_t = \sqrt{\gamma} \beta_t dt + dW_t,
\end{equation}
where $\beta_s = \beta_t(\alpha,Y^t)$ is a $\mathcal{A}\otimes \cB_{s-}(C)$ measurable function such that $\int_0^T \bE \beta_s^2 ds <\infty$, we have
\begin{equation}
I(\alpha; Y^T) = \int_0^T \bE \ell_\Gscr(\beta_s, \hat{\beta}^{P,C}_s(\gamma))ds.
\end{equation}
\end{corollary}

\subsubsection{Multivariate Point Process Channel}

The multivariate point process channel model is a generalization of the Poisson channel model, where the the output process may have various jump sizes. The $(\tau_n)$-uniqueness property is also satisfied in this situation~\cite{Jacod--Shiryaev1987}. To be precise, under SNR $\gamma$, the output process $(Y_t, \bF, P)$ in the absence of input is a L\'evy process with the following representation:
\begin{equation}
Y_t = \int_0^t \int_{|z|<1} z (d\mu - \gamma \nu(dz)ds) + \int_0^t \int_{|z|\geq 1} z d\mu.
\end{equation}

For encodings $\lambda_{s,z} = \lambda_{s,z}(\alpha, Y^{s-}) \geq 0$, the new output process $(Y_t, \bF, P)$ could be represented as
\begin{equation}\label{eqn.mppmodel}
Y_t = \int_0^t \int_{|z|< 1} \gamma z (\lambda_{s,z}-1)\nu(dz)ds + \int_0^t \int_{|z|< 1} z (d\mu - \gamma \lambda_{s,z} \nu(dz)ds) + \int_0^t \int_{|z|\geq 1} z d\mu.
\end{equation}

We have the following representation for the mutual information $I(\alpha; Y^T)$ for the multivariate point process channel~\cite[Thm. 19.11]{Liptser--Shiryaev2001}. 

\begin{corollary}\label{cor.mppmodel}
Under channel model~(\ref{eqn.mppmodel}), if 
\begin{align}
\int_0^T \int_{\mathbb{R}_0} \mathbb{E} \left[ \ell_{\mathcal{P}}(\lambda_{s,z},\hat{\lambda}^{P,C}_{s,z}(\gamma) ) + 2 \lambda_{s,z} \right]\nu(dz)ds <\infty, 
\end{align}
then,
\begin{equation}
I(\alpha;Y^T) = \int_0^T \int_{\bR_0} \bE \ell_\Pscr(\lambda_{s,z}, \hat{\lambda}^{P,C}_{s,z}(\gamma))\nu(dz)ds,
\end{equation}
where $\hat{\lambda}_{s,z}^{P,C}(\gamma) = \mathbb{E}_P[\lambda_{s,z}| \cF_{s-}^Y]$, $\ell_{\mathcal{P}}(x,y) = x\ln \frac{x}{y}-x+y$.
\end{corollary}

\subsection{Doob--Meyer decomposition of a class of sub-martingales}

Since $-\log(\cdot)$ is a convex function, it is clear that that
\begin{equation}\label{eqn.ratiopoint}
\log \frac{dP_{Y^t}}{dQ_{Y^t}} 
\end{equation}
is a $P$-sub-martingale. Since we know under mild conditions, any sub-martingale can be decomposed uniquely into the sum of a predictable non-decreasing process and a local martingale~\cite[Chap. 5]{medvegyev2007stochastic}, i.e., the Doob--Meyer decomposition, it arises as a natural question to find the Doob--Meyer decomposition of (\ref{eqn.ratiopoint}). Although in general it is a hard task to obtain explicit expressions for the  Doob--Meyer decomposition of sub-martingales, we show in this case it has an elegant answer, with implications for relations between information and estimation. In particular, we observe that the expectation of the predictable non-decreasing process is precisely the filtering error. 

\begin{theorem}\label{thm. SMG relative information}
Under Assumption~\ref{ass.inputdefinition}, we have 
\begin{equation}
\log \frac{dP_{Y^t}}{dQ_{Y^t}} = A_t + M_t,
\end{equation}
where
\begin{align}
A_t & = \gamma \int_0^t \ell_{\mathcal{G}}(\hat{\beta}_s^{P,C}(\gamma),\hat{\beta}_s^{Q,C}(\gamma))ds
+ \gamma \int_0^t \int_{\bR_0} \ell_\Pscr(\hat{\lambda}_{s,z}^{P,C}(\gamma),\hat{\lambda}_{s,z}^{Q,C}(\gamma)) \nu(dz)ds,  \\ 
M_t & = \sqrt{\gamma} \int_0^t \left( \hat{\beta}_s^{P,C}(\gamma) - \hat{\beta}_s^{Q,C}(\gamma)\right)(dW_s - \sqrt{\gamma} \hat{\beta}_s^{P,C}(\gamma) ds) \nonumber \\
& \quad + \int_0^t \int_{\bR_0} \ln \frac{\hat{\lambda}_{s,z}^{P,C}(\gamma)}{\hat{\lambda}_{s,z}^{Q,C}(\gamma)}  (d\mu - \gamma \hat{\lambda}_{s,z}^{P,C}(\gamma) \nu(dz)ds),
\end{align}
where $\hat{\beta}_s^{P,C}(\gamma) = \mathbb{E}_P[\beta_s|\cF_{s-}^Y]$, $\hat{\lambda}_{s,z}^{P,C}(\gamma) = \mathbb{E}_P[\lambda_{s,z}| \cF_{s-}^Y]$, $\ell_{\mathcal{G}}(x,y) = \frac{1}{2}(x-y)^2, \ell_{\mathcal{P}}(x,y) = x\ln \frac{x}{y}-x+y$. Here the process $A_t$ is the predictable non-decreasing process, and $M_t$ is the local martingale process. 
\end{theorem}

Specializing Theorem~\ref{thm. SMG relative information} to the case of $P$ being deterministic and $Q = P$, we obtain the following Doob--Meyer decomposition for the information density process 
\begin{equation}
\log \frac{dP_{Y^t|\alpha}}{dP_{Y^t}}. 
\end{equation}

\begin{theorem} \label{thm. SMG information density}
Under Assumption~\ref{ass.inputdefinition}, we have 
\begin{equation}
\log \frac{dP_{Y^t|\alpha}}{dP_{Y^t}} = A_t + M_t,
\end{equation}
where
\begin{align}
A_t & = \gamma \int_0^t \ell_{\mathcal{G}}(\beta_s ,\hat{\beta}_s^{P,C}(\gamma)) ds + \gamma \int_0^t \int_{\bR_0} \ell_\Pscr(\lambda_{s,z}, \hat{\lambda}_{s,z}^{P,C}(\gamma)) \nu(dz)ds,  \\ 
M_t & = \sqrt{\gamma} \int_0^t (\beta_s - \hat{\beta}_s^{P,C}(\gamma))(dW_s - \sqrt{\gamma} \beta_s ds) + \int_0^t \int_{\bR_0} \ln \frac{\lambda_{s,z}}{\hat{\lambda}_{s,z}^{P,C}(\gamma))} (d\mu - \gamma \lambda_{s,z} \nu(dz)ds),
\end{align}
where $\hat{\beta}_s^{P,C}(\gamma) = \mathbb{E}_P[\beta_s|\cF_{s-}^Y]$, $\hat{\lambda}_{s,z}^{P,C}(\gamma) = \mathbb{E}_P[\lambda_{s,z}| \cF_{s-}^Y]$, $\ell_{\mathcal{G}}(x,y) = \frac{1}{2}(x-y)^2, \ell_{\mathcal{P}}(x,y) = x\ln \frac{x}{y}-x+y$. Here the process $A_t$ is the predictable non-decreasing process, and $M_t$ is the local martingale process. 
\end{theorem}

\section{Proofs}\label{sec: SMG proofs}

Our focus would be to establish the Doob--Meyer decomposition for the $P$-sub-martingale $\log \frac{dP_{Y^t}}{dQ_{Y^t}}$ (Theorem~\ref{thm. SMG relative information}), from which the rest of our results will follow. Recall that at SNR level $\gamma$, in the absence of input signal, the output process $(Y_t, \bF, P_0)$ of a semi-martingale channel at SNR $\gamma$ is a L\'evy process with the following representation:
\begin{equation}\label{eqn.levyitoSMG}
Y_t = W_t + \int_0^t \int_{|z|< 1} z (d\mu - \gamma \nu(dz)ds) + \int_0^t \int_{|z|\geq 1} zd \mu,
\end{equation}
where $W_t$ is a standard Brownian motion, $\mu(dt,dz)$ is a Poisson random measure on $[0,T]\times \bR_0$, independent of $W^T$.  

Introduce the non-negative process $(L_t, \bF, P)$, where $\mathbb{R}_0 = \mathbb{R} \backslash \{0\}$, as
\begin{equation}\label{eqn.ratioprocess}
L_t = e^{\sqrt{\gamma} \int_0^t \beta_s dW_s - \frac{\gamma}{2} \int_0^t \beta_s^2 ds + \int_0^t \int_{\bR_0} \left [ \ln \lambda_{s,z} d\mu - \gamma (\lambda_{s,z}-1)\nu(dz)ds \right]}.
\end{equation}

We have the following It$\hat{\mathrm{o}}$'s formula for general semimartingales:
\begin{lemma}\cite[Thm. 6.46]{medvegyev2007stochastic}\label{lemma.itosemimg}
If $\{Z(t):t\geq 0\}$ is a semimartingale and $f(x) \in C^2(\mathbb{R})$, then
\begin{align}
f(Z(t)) - f(Z(0)) & = \int_0^t f'(Z_{-})dZ + \frac{1}{2} \int_0^t f''(Z_{-}) d[Z]^c + \sum_{0<s\leq t} \left( f(Z(s)) -f(Z(s-)) - f'(Z(s-))\Delta Z(s) \right),
\end{align}
where the process $[Z]^c_t$ is the quadratic variation process of the continuous part of the semimartingale $Z(t)$, $\Delta Z(s) = Z(s) - Z(s-)$, and $Z(s-) = \lim_{u\to s-} Z(u)$. 
\end{lemma}

Applying Lemma~\ref{lemma.itosemimg} with $f(t) = e^t$, defining $D_t = \int_0^t \int_{\bR_0} \left [ \ln \lambda_{s,z} d\mu - \gamma (\lambda_{s,z}-1)\nu(dz)ds \right]$, we get the following representation of the stochastic process $L_t$:
\begin{align}
L_t & = 1 + \int_0^t L_{s-} dZ(t) + \frac{1}{2} \int_0^t L_{s-} \gamma \beta_s^2 ds + \sum_{0<s\leq t} f(Z(s-)) \left( \frac{f(Z(s))}{f(Z(s-))} - 1 - \Delta Z(s) \right) \\
& = 1 + \int_0^t \sqrt{\gamma} \beta_s L_{s-} dW_s + \int_0^t L_{s-} dD_s + \sum_{0<s\leq t} L_{s-} \left( e^{\Delta Z(s)} - 1 - \Delta Z(s) \right) \\
& = 1 + \int_0^t \sqrt{\gamma} \beta_s L_{s-} dW_s + \int_0^t L_{s-} dD_s + \sum_{0<s\leq t} \int_{\mathbb{R}_0} L_{s-} \left( e^{\ln \lambda_{s,z}} - 1 - \ln \lambda_{s,z} \right) \nu(ds,dz) \\
& = 1 + \int_{0}^t L_{s-}dM_s,
\end{align} 
where 
\begin{align}
M_t & = \int_0^t \sqrt{\gamma} \beta_s dW_s + \int_0^t \int_{\mathbb{R}_0} \left(  \ln \lambda_{s,z} \mu(ds,dz) - \gamma(\lambda_{s,z}-1) \nu(dz)ds + (\lambda_{s,z} - 1 - \ln \lambda_{s,z} )\mu(ds,dz) \right) \\
& = \int_0^t \sqrt{\gamma} \beta_s dW_s + \int_0^t \int_{\mathbb{R}_0} 
(\lambda_{s,z}-1) (\mu(ds,dz) - \gamma \nu(dz)ds). 
\end{align}

It follows from~\cite[Thm. 12]{Kabanov--Liptser--Shiryaev1978-1} that if there exists a constant $V>0$ such that
\begin{equation}
\int_0^T \beta_s^2 ds + \int_0^T \int_{\bR_0} (1-\sqrt{\lambda_{s,z}})^2 \nu(dz)ds \leq V\quad P-a.s.
\end{equation}
then, $\{L_t: 0\leq t\leq T\}$ is a uniformly integrable martingale. It is guaranteed by Assumption~\ref{ass.inputdefinition}. Construct another probability measure $P$ on $\mathbb{F}$ defined as
\begin{align}
\frac{d P_{|\mathcal{F}_t}}{dP_{0|\mathcal{F}_t}} & = L_t,
\end{align}

It follows from~\cite[Corollary, pg. 663]{Kabanov--Liptser--Shiryaev1978-1} that under measure $P$, the process $Y_t$ is still a semi-martingale with predictable characteristics 
\begin{equation}
(\sqrt{\gamma}\int_0^t \beta_s ds + \int_0^t \int_{|z|\leq 1} \gamma z (\lambda_{s,z}-1)\nu(dz)ds, t, \gamma \lambda_{t,z} \nu(dz)dt),
\end{equation}
which is exactly what we specified in the definition of the semi-martingale channel in (\ref{eqn.pcnew}). Since we have assumed that the measure $P$ has $(\tau_n)$-uniqueness property, if we take $\tau_n \equiv T$, we know that $P$ is the probability measure governing the output of the semi-martingale channel with input signals $\beta_s$ and $\lambda_{s,z}$. 

It follows from~\cite[Chap. 4, Sec. 6, Thm. 5]{Liptser--Shiryaev1989} that the semi-martingale $(Y_t, \bF, P)$ is still a semi-martingale under the reduced filtration $\cF_t^Y = \sigma\{Y_s: s\leq t\}$. Under the filtration $\mathcal{F}_t^Y$, combining with Assumption~\ref{ass.inputdefinition} the predictable characteristics of process $Y_t$ would change to 
\begin{equation}
(\sqrt{\gamma}\int_0^t \hat{\beta}_s^{P,C}(\gamma) ds + \int_0^t \int_{|z|\leq 1} \gamma z (\hat{\lambda}_{s,z}^{P,C}(\gamma)-1)\nu(dz)ds, t, \gamma \hat{\lambda}_{s,z}^{P,C}(\gamma)\nu(dz)dt),
\end{equation}
where $\hat{\beta}_s^{P,C}(\gamma) = \mathbb{E}_P[\beta_s|\cF_{s-}^Y]$, $\hat{\lambda}_{s,z}^{P,C}(\gamma) = \mathbb{E}_P[\lambda_{s,z}| \cF_{s-}^Y]$. 

It follows from the convexity of $x^2$ and $(1-\sqrt{x})^2$ on $\mathbb{R}$ and $\mathbb{R}_+$, respectively, that 
\begin{align}
\int_0^T [\hat{\beta}_s^{P,C}(\gamma)]^2 ds + \int_0^T \int_{\bR_0} \left (1-\sqrt{\hat{\lambda}_{s,z}^{P,C}(\gamma)} \right )^2   \nu(dz)ds & \leq \mathbb{E}_P \left[ \int_0^T \beta_s^2 ds + \int_0^T \int_{\bR_0} \left( 1-\sqrt{\lambda_{s,z}} \right)^2 \nu(dz)ds \Bigg | \mathcal{F}_s^Y \right]. 
\end{align}
It then follows from the fact that for any random variable $X$ and constant $V$, $X\leq V$ almost surely implies that $\mathbb{E}[X|\mathcal{F}]\leq V$ almost surely, that 
\begin{equation}
\int_0^T [\hat{\beta}_s^{P,C}(\gamma)]^2 ds + \int_0^T \int_{\bR_0} \left (1-\sqrt{\hat{\lambda}_{s,z}^{P,C}(\gamma)} \right )^2   \nu(dz)ds \leq V, \quad P-a.s.
\end{equation}

Hence, 
\begin{equation}
\bar{L}_t^P = e^{\sqrt{\gamma} \int_0^t\hat{\beta}_s^{P,C}(\gamma) dW_s - \frac{\gamma}{2} \int_0^t [\hat{\beta}_s^{P,C}(\gamma)]^2 ds + \int_0^t \int_{\bR_0} \left [ \ln \hat{\lambda}_{s,z}^{P,C}(\gamma) d\mu - \gamma (\hat{\lambda}_{s,z}^{P,C}(\gamma)-1)\nu(dz)ds \right]}
\end{equation}
is a uniformly integrable martingale~\cite[Thm. 12]{Kabanov--Liptser--Shiryaev1978-1}. Using similar arguments as above and applying the $(\tau_n)$-uniqueness property, we know that
\begin{align}
\bar{L}_t^P & = \frac{dP_{|\mathcal{F}_t^Y}}{dP_{0|\mathcal{F}_t^Y}}. 
\end{align}

Analogously, if the input signals follow distribution $Q$, we can use similar arguments to construct the likelihood ratio process $\bar{L}_t^Q$. Hence, 
\begin{align}
\log \frac{dP_{|\mathcal{F}_t^Y}}{dQ_{|\mathcal{F}_t^Y}} & = \log \frac{dP_{|\mathcal{F}_t^Y}}{dP_{0|\mathcal{F}_t^Y}} - \log \frac{dQ_{|\mathcal{F}_t^Y}}{dP_{0|\mathcal{F}_t^Y}} \\
& = \sqrt{\gamma} \int_0^t \hat{\beta}_s^{P,C}(\gamma) dW_s - \frac{\gamma}{2} \int_0^t [\hat{\beta}_s^{P,C}(\gamma)]^2 ds + \int_0^t \int_{\bR_0} \left [ \ln \hat{\lambda}_{s,z}^{P,C}(\gamma) d\mu - \gamma (\hat{\lambda}_{s,z}^{P,C}(\gamma)-1)\nu(dz)ds \right] \nonumber \\
& \quad - \left( \sqrt{\gamma} \int_0^t\hat{\beta}_s^{Q,C}(\gamma) dW_s - \frac{\gamma}{2} \int_0^t [\hat{\beta}_s^{Q,C}(\gamma)]^2 ds + \int_0^t \int_{\bR_0} \left [ \ln \hat{\lambda}_{s,z}^{Q,C}(\gamma) d\mu - \gamma (\hat{\lambda}_{s,z}^{Q,C}(\gamma)-1)\nu(dz)ds \right] \right) \\
& = \sqrt{\gamma} \int_0^t ( \hat{\beta}_s^{P,C}(\gamma) - \hat{\beta}_s^{Q,C}(\gamma) ) (dW_s - \sqrt{\gamma} \hat{\beta}_s^{P,C}(\gamma) ds) + \frac{\gamma}{2}\int_0^t \left( \hat{\beta}_s^{P,C}(\gamma) - \hat{\beta}_s^{Q,C}(\gamma) \right)^2 ds \nonumber \\
& \quad + \int_0^t \int_{\mathbb{R}_0} \ln \frac{\hat{\lambda}_{s,z}^{P,C}(\gamma)}{\hat{\lambda}_{s,z}^{Q,C}(\gamma)} (d\mu - \hat{\lambda}_{s,z}^{P,C}(\gamma) \nu(dz)ds) + \gamma \int_0^t \int_{\mathbb{R}_0} \ell_{\mathcal{P}} (\hat{\lambda}_{s,z}^{P,C}(\gamma), \hat{\lambda}_{s,z}^{Q,C}(\gamma) )\nu(dz)ds. 
\end{align}

The proof of Theorem~\ref{thm. SMG relative information} is now complete. To obtain the representations of relative entropy, it suffices to take expectations of $\log \frac{dP_{|\mathcal{F}_t^Y}}{dQ_{|\mathcal{F}_t^Y}}$ with respect to the measure induced by $P$. Indeed, it follows from the results of~\cite[Chap. 4, Sec. 6, Thm. 5]{Liptser--Shiryaev1989} that $W_t -  \sqrt{\gamma} \int_0^t \hat{\beta}_s^{P,C}(\gamma) ds$ is a standard Brownian motion under filtration $\cF_t^Y$ with probability measure $P$. Since we have assumed $\int_0^T \mathbb{E}  \left( \hat{\beta}_s^{P,C}(\gamma) - \hat{\beta}_s^{Q,C}(\gamma) \right)^2 ds < \infty$, it follows from~\cite[Chap. 5.4]{Liptser--Shiryaev2000} that
\begin{equation}
\bE  \left[ \sqrt{\gamma} \int_0^t ( \hat{\beta}_s^{P,C}(\gamma) - \hat{\beta}_s^{Q,C}(\gamma) ) (dW_s - \sqrt{\gamma} \hat{\beta}_s^{P,C}(\gamma) ds)  \right ]  = 0.
\end{equation}
Since we have assumed 
\begin{align}
\int_0^t \int_{\mathbb{R}_0} \mathbb{E} \lambda_{s,z} \left| \ln \frac{\hat{\lambda}_{s,z}^{P,C}(\gamma)}{\hat{\lambda}_{s,z}^{Q,C}(\gamma)} \right | \nu(dz)ds < \infty,
\end{align}
it follows from~\cite[Thm. 18.7]{Liptser--Shiryaev2001} that 
\begin{align}
\mathbb{E} \left[  \int_0^t \int_{\mathbb{R}_0} \ln \frac{\hat{\lambda}_{s,z}^{P,C}(\gamma)}{\hat{\lambda}_{s,z}^{Q,C}(\gamma)} (d\mu - \hat{\lambda}_{s,z}^{P,C}(\gamma) \nu(dz)ds) \right] = 0. 
\end{align}
Theorem~\ref{thm.relative} is proved. Theorem~\ref{thm.mutual} can be proved in a similar fashion.

We now provide a proof sketch for Corollary~\ref{cor.mppmodel}. It was shown in~\cite[Sec. 12]{Kabanov--Liptser--Shiryaev1978-2} that 
\begin{align}
\mathbb{E} \int_0^T \int_{\mathbb{R}_0} ( 1- \sqrt{\lambda_{s,z}})^2 \nu(dz) ds < \infty
\end{align}
implies that $P \ll P_0$, where $P_0$ is the probability measure on the output process without inputs, and $P$ is the measure corresponding to inputs $\lambda_{s,z}$. Following similar arguments as in~\cite[Thm. 19.11]{Liptser--Shiryaev2001} and noting that $x|\ln x/y| \leq \ell_{\mathcal{P}}(x,y) + x + y$, and $(1-\sqrt{x})^2 \leq C \ell_{\mathcal{P}}(x,1)$ for some constant $C>0$, Corollary~\ref{cor.mppmodel} is proved.

\section{Concluding remarks} \label{sec: SMG summary}

At the face of it, the output stochastic process of the semi-martingale channel seems to be a simple combination of a `continuous' process and a `pure jump' process. Indeed, one can separate these two processes at the receiver perfectly. However, it is important to note that the inputs may causally depend on past outputs of both the continuous part and the pure jump part! As the careful reader will note, the conditional expectations in Theorem~\ref{thm.mutual} are taken with respect to the entire history (including the continuous part and discontinuous part) of $Y$, which is not the same as treating the continuous and discontinuous outputs separately.

Relations between information and estimation are, at their core intimately related to absolute continuity and singularity of probability measures in functional spaces, which enables explicit calculations of the most basic likelihood ratios, such as the information density and the relative information. Shiryaev~\cite{Shiryaev1978absolute} presented a framework of the general theory of absolute continuity and singularity of probability measures, which gives us a good understanding of the representation of likelihood ratios for random sequences, processes with independent increments, semi-martingales with a Gaussian martingale component, multivariate point processes, Markov processes and processes with a countable number of states, and the general semi-martingales~\cite{Kabanov--Liptser--Shiryaev1978-2}. This rich theory essentially implies that if the output of a channel is of the types above, and a natural SNR parameter can be defined, one may hope to get a general and meaningful relationship between measures of information and estimation. These two constraints essentially make the semi-martingale channels the largest class of channels that admit information-estimation relationships fully paralleling what exist for the Gaussian and Poisson channels. However, we note that the likelihood ratio characterization for semi-martingales is challenging, and much stronger conditions are needed to represent these likelihood ratios. This is precisely the reason why Theorems~\ref{thm.mutual},\ref{thm.relative} require strong (bounded a.s.) conditions, and special cases of semi-martingale channels can be dealt with under much weaker conditions on the channel input, as evident in Section~\ref{sec.special}. 

\bibliographystyle{IEEEtran}
\bibliography{references}

\newcommand{\noopsort}[1]{}
\begin{thebibliography}{10}
\providecommand{\url}[1]{#1}
\csname url@samestyle\endcsname
\providecommand{\newblock}{\relax}
\providecommand{\bibinfo}[2]{#2}
\providecommand{\BIBentrySTDinterwordspacing}{\spaceskip=0pt\relax}
\providecommand{\BIBentryALTinterwordstretchfactor}{4}
\providecommand{\BIBentryALTinterwordspacing}{\spaceskip=\fontdimen2\font plus
\BIBentryALTinterwordstretchfactor\fontdimen3\font minus
  \fontdimen4\font\relax}
\providecommand{\BIBforeignlanguage}[2]{{%
\expandafter\ifx\csname l@#1\endcsname\relax
\typeout{** WARNING: IEEEtran.bst: No hyphenation pattern has been}%
\typeout{** loaded for the language `#1'. Using the pattern for}%
\typeout{** the default language instead.}%
\else
\language=\csname l@#1\endcsname
\fi
#2}}
\providecommand{\BIBdecl}{\relax}
\BIBdecl

\bibitem{Shannon1949}
C.~Shannon, ``Communication in the presence of noise,'' \emph{Proceedings of
  the IEEE}, vol.~86, no.~2, pp. 447--457, Feb 1998.

\bibitem{Duncan1970}
T.~E. Duncan, ``On the calculation of mutual information,'' \emph{SIAM Journal
  on Applied Mathematics}, vol.~19, no.~1, pp. 215--220, 1970.

\bibitem{Kadota--Zakai--Ziv1971}
T.~Kadota, M.~Zakai, and J.~Ziv, ``Mutual information of the white {Gaussian}
  channel with and without feedback,'' \emph{Information Theory, IEEE
  Transactions on}, vol.~17, no.~4, pp. 368--371, 1971.

\bibitem{Kabanov1978}
Y.~M. Kabanov, ``The capacity of a channel of the {Poisson} type,''
  \emph{Theory of Probability \& Its Applications}, vol.~23, no.~1, pp.
  143--147, 1978.

\bibitem{Atar--Weissman2012}
R.~Atar and T.~Weissman, ``Mutual information, relative entropy, and estimation
  in the {Poisson} channel,'' \emph{Information Theory, IEEE Transactions on},
  vol.~58, no.~3, pp. 1302--1318, 2012.

\bibitem{Liptser--Shiryaev2001}
R.~Liptser and A.~N. Shiryaev, \emph{Statistics of Random Processes II:
  Applications, 2nd ed.}\hskip 1em plus 0.5em minus 0.4em\relax
  Springer-Verlag, 2001.

\bibitem{Guo--Shamai--Verdu2008}
D.~Guo, S.~Shamai, and S.~Verd{\'u}, ``Mutual information and conditional mean
  estimation in {Poisson} channels,'' \emph{{Information Theory, IEEE
  Transactions on}}, vol.~54, no.~5, pp. 1837--1849, 2008.

\bibitem{Duncan2010}
T.~Duncan, ``Mutual information for stochastic signals and {L{\'{e}}vy}
  processes,'' \emph{Information Theory, IEEE Transactions on}, vol.~56, no.~1,
  pp. 18--24, Jan 2010.

\bibitem{Johnson2013}
\BIBentryALTinterwordspacing
O.~Johnson, ``A de bruijn identity for symmetric stable laws,'' \emph{CoRR},
  vol. abs/1310.2045, 2013. [Online]. Available:
  \url{http://arxiv.org/abs/1310.2045}
\BIBentrySTDinterwordspacing

\bibitem{Grigelionis1974}
B.~Grigelionis, ``Mutual information for locally infinitely divisible random
  processes,'' \emph{Lithuanian Mathematical Journal}, vol.~14, no.~1, pp.
  1--6, 1974.

\bibitem{Guo--Shamai--Verdu2005}
D.~Guo, S.~Shamai, and S.~Verd{\'u}, ``Mutual information and minimum
  mean-square error in {Gaussian} channels,'' \emph{Information Theory, IEEE
  Transactions on}, vol.~51, no.~4, pp. 1261--1282, 2005.

\bibitem{Verdu2006}
S.~Verd{\'u} and D.~Guo, ``A simple proof of the entropy-power inequality,''
  \emph{IEEE Transactions on Information Theory}, vol.~52, no.~5, pp.
  2165--2166, 2006.

\bibitem{Tulino--Verdu2006}
A.~M. Tulino and S.~Verd{\'u}, ``Monotonic decrease of the non-{Gaussianness}
  of the sum of independent random variables: A simple proof,''
  \emph{Information Theory, IEEE Transactions on}, vol.~52, no.~9, pp.
  4295--4297, 2006.

\bibitem{Weissman2010}
T.~Weissman, ``The relationship between causal and non-causal mismatched
  estimation in continuous-time {AWGN} channels,'' \emph{Information Theory,
  IEEE Transactions on}, vol.~56, no.~9, pp. 4256--4273, 2010.

\bibitem{Venkat--Weissman2012}
K.~Venkat and T.~Weissman, ``Pointwise relations between information and
  estimation in {Gaussian} noise,'' \emph{Information Theory, IEEE Transactions
  on}, vol.~58, no.~10, pp. 6264--6281, 2012.

\bibitem{Jiao--Venkat--Weissman2013}
J.~Jiao, K.~Venkat, and T.~Weissman, ``Pointwise relations between information
  and estimation in the {Poisson} channel,'' in \emph{Information Theory
  Proceedings (ISIT), 2013 IEEE International Symposium on}.\hskip 1em plus
  0.5em minus 0.4em\relax IEEE, 2013, pp. 449--453.

\bibitem{Jiao--Venkat--Weissman_ISIT2014}
------, ``Relations between information and estimation in scalar {L{\'{e}}vy}
  channels,'' in \emph{Information Theory (ISIT), 2014 IEEE International
  Symposium on}, June 2014, pp. 2212--2216.

\bibitem{Jiao--Venkat--Weissman_TIT2014}
------, ``Relations between information and estimation in discrete-time
  {L}\'evy channels,'' \emph{to appear in IEEE Transactions on Information
  Theory}, 2017.

\bibitem{medvegyev2007stochastic}
P.~Medvegyev, \emph{Stochastic integration theory}.\hskip 1em plus 0.5em minus
  0.4em\relax Oxford University Press on Demand, 2007, no.~14.

\bibitem{Liptser--Shiryaev1989}
R.~Liptser and A.~Shiryaev, \emph{Theory of martingales}.\hskip 1em plus 0.5em
  minus 0.4em\relax Kluwer Academic Publishers (Dordrecht and Boston), 1989,
  vol.~49.

\bibitem{Rao1999}
B.~S.~P. Rao, \emph{Semimartingales and their statistical inference}.\hskip 1em
  plus 0.5em minus 0.4em\relax CRC Press, 1999, vol.~83.

\bibitem{Jacod--Shiryaev1987}
J.~Jacod and A.~N. Shiryaev, \emph{Limit theorems for stochastic
  processes}.\hskip 1em plus 0.5em minus 0.4em\relax Springer-Verlag Berlin,
  1987, vol. 288.

\bibitem{Kabanov--Liptser--Shiryaev1978-2}
Y.~M. Kabanov, R.~S. Liptser, and A.~N. Shiryaev, ``Absolute continuity and
  singularity of locally absolutely continuous probability distributions. ii,''
  \emph{Matematicheskii Sbornik}, vol. 150, no.~1, pp. 32--61, 1979.

\bibitem{Jacod--Memin1976}
J.~Jacod and J.~Memin, ``Caract{\'e}ristiques locales et conditions de
  continuit{\'e} absolue pour les semi-martingales,'' \emph{Probability Theory
  and Related Fields}, vol.~35, no.~1, pp. 1--37, 1976.

\bibitem{Skorokhod1982}
A.~V. Skorokhod, \emph{Studies in the theory of random processes}.\hskip 1em
  plus 0.5em minus 0.4em\relax Dover New York, 1982.

\bibitem{Kabanov--Liptser--Shiryaev1980}
Y.~M. Kabanov, R.~S. Liptser, and A.~Shiryayev, ``On absolute continuity of
  probability measures for {M}arkov-{I}to processes,'' in \emph{Stochastic
  Differential Systems Filtering and Control}.\hskip 1em plus 0.5em minus
  0.4em\relax Springer, 1980, pp. 114--128.

\bibitem{Sato1999}
K.-i. Sato, \emph{L{\'e}vy processes and infinitely divisible
  distributions}.\hskip 1em plus 0.5em minus 0.4em\relax Cambridge university
  press, 1999.

\bibitem{Kuchler--Sorensen1997}
U.~K{\"u}chler and M.~S{\o}rensen, ``Exponential families of stochastic
  processes. 1997.''

\bibitem{Liptser--Shiryaev2000}
R.~Liptser and A.~N. Shiryaev, \emph{Statistics of Random Processes I: General
  Theory, 2nd ed.}\hskip 1em plus 0.5em minus 0.4em\relax Springer-Verlag,
  2000.

\bibitem{Kabanov--Liptser--Shiryaev1978-1}
Y.~M. Kabanov, R.~S. Liptser, and A.~N. Shiryaev, ``Absolute continuity and
  singularity of locally absolutely continuous probability distributions. i,''
  \emph{Matematicheskii Sbornik}, vol. 149, no.~3, pp. 364--415, 1978.

\bibitem{Shiryaev1978absolute}
A.~Shiryaev, ``Absolute continuity and singularity of probability measures in
  functional spaces,'' in \emph{Proceedings of the International Congress of
  Mathematicians, Helsinki}, 1978, pp. 209--225.

\end{thebibliography}

\end{document}